RESEARCH ARTICLE

# Peak-Nadir Encoding for Efficient CGM Data Compression and High-Fidelity Reconstruction

Running title:

Efficient CGM Data Compression

Authors:

Clara Bender MSc PhD[1], Line Davidsen MD[2,3],
Søren Schou Olesen MD PhD[2,3], Simon Lebech Cichosz MSc BMEI PhD[1]

Author Affiliations:

[1]Department of Health Science and Technology, Aalborg University, Denmark. [2]Department of Clinical Medicine, Aalborg University Hospital, Aalborg, Denmark. [3]Centre for Pancreatic Diseases and Mech-Sense, Department of Gastroenterology and Hepatology, Aalborg University Hospital, Aalborg, Denmark.

Corresponding author:

Simon Cichosz, simcich@hst.aau.dk, Postal address: Selma Lagerløfs Vej 249, 12-02-048, 9260 Gistrup, Danmark, Phone: (+45) 9940 2020; Fax: (+45) 9815 4008. ORCID: 0000-0002-3484-7571

Figures: 5
Tables: 2
Words: 2801

Abbreviations: Conditional Generative Adversarial Network (CGAN); Continuous Glucose Monitoring (CGM); Coefficient of Variation (CV); Mean Absolute Error (MAE); Mean Amplitude of Glycemic Excursions (MAGE); Peaks & Nadirs (PN); Peaks, Nadirs, and Support Points (PN+); Piecewise Cubic Hermite Interpolating Polynomial (PCHIP); Quantification of Continuous Glucose Monitoring (QoCGM); Standard Deviation (Std); Time-Above-Range (TAR); Time-Below Range (TBR); Time-In-Range (TIR); Time-In-Tight-Range (TITR)

# Abstract

**Aim/background**: Continuous glucose monitoring (CGM) generates dense time-series data, posing challenges for efficient storage, transmission, and analysis. This study evaluates novel encoding strategies that reduce CGM profiles to a compact set of landmark points while maintaining fidelity in reconstructed signals and derived glycemic metrics.

**Methods**: We utilized two complementary CGM datasets, synthetic data generated via a Conditional Generative Adversarial Network (CGAN) and real-world measurements from a randomized crossover trial, to develop and validate three encoding approaches: (1) Peaks & Nadirs (PN), (2) Peaks, Nadirs, and Support Points (PN+), and (3) Uniform Downsampling. Each method compresses CGM profiles by selecting key timestamps and glucose values, followed by signal reconstruction via interpolation. Performance was assessed using compression ratio, mean absolute error (MAE), and $R^2$ between original and reconstructed clinically relevant CGM-derived metrics. Statistical analyses evaluated the preservation of clinically relevant glucose features.

**Results**: Across varying compression settings, PN+ consistently outperformed PN and downsampling, achieving the highest $R^2$ and lowest MAE. At a compression ratio of 13 (22 landmark points per 24-hour profile), PN+ reduced MAE by a factor of 3.6 compared to downsampling (0.77 vs. 2.75), with notable improvements in metrics sensitive to glucose excursions. Encoding and decoding required an average of 0.13 seconds per profile. Validation on real-world data confirmed these trends.

**Conclusions**: The proposed PN+ method produces a compact CGM representation that retains critical glycemic dynamics while discarding redundant portions of the profiles. The CGM signal can be reconstructed with high precision from the encoding representation.

# Introduction

Continuous glucose monitoring (CGM) has become a cornerstone of modern diabetes care, yielding a time series of glucose measurements every few minutes [1–3]. This high-resolution data stream has revolutionized both clinical management and research, enabling advanced evaluation of therapies, classification of patient subgroups, data-driven prediction of glucose dynamics, and discovery of glycemic patterns via machine learning [4–9]. For example, the adoption of CGM in type 1 diabetes has surged in recent years, reflecting its importance for decision support and personalized treatment [10]. However, the richness of CGM data and other biomedical data also poses challenges [11]: large numbers of daily readings per patient produce very large datasets that must be stored, shared, and processed efficiently.

Storing and interoperating on such high-frequency data can be prohibitive. As Jacobsson *et al.* [11] state, modern healthcare systems face an "ever-increasing need for retrieving, storing, and managing the large amount of biomedical signal data generated," and common data exchange standards like HL7 FHIR are inefficient when used as a long-term storage format. Naively saving CGM streams as text-based FHIR records or CSV files incurs high overhead and wastes storage. These inefficiencies hinder scalable CGM data integration and slow down analysis. Together, these challenges highlight the need to explore compression or compact encoding methods that reduce data size while preserving critical clinical information [11].

Efficient encoding and compression are also key to secure data sharing and interoperability. In emerging healthcare architectures, technologies like blockchain and internet-of-things require lean data formats to function on a scale [12,13]. For instance, blockchain-based platforms for diabetes care have been proposed to improve data sharing and patient-centric control [14], but the limited throughput of distributed ledgers means that raw CGM time series cannot be stored on-chain without preprocessing. By compressing and encoding glucose traces into concise, standardized representations, data can be embedded in interoperable payloads or off-chain repositories and exchanged securely between providers and patients.

At the same time, CGM analysis increasingly relies on AI and deep learning. Neural networks for glucose forecasting and pattern recognition often require well-structured, fixed-dimensional input and benefit from reduced noise. Textual encoding of CGM traces has been shown to greatly aid such analytics. For example, Igbe *et al.* (2025) [15] demonstrated that mapping CGM profiles to reduced-alphabet "strings" could enable powerful search and classification, as well as support predictive modeling, anomaly detection, and even generative AI applications. In clinical practice, alphabetic encoding could facilitate data indexing, reduce search time for glycemic patterns, and reduce computational complexity by leveraging text-based data structures and machine learning pipelines. Such encoded biomedical data can be fed directly into deep learning models to improve the speed and accuracy of analysis [16,17].

Collectively, challenges such as data volume and storage, interoperability, secure sharing, and AI readiness motivate the need for an efficient compression and encoding/decoding framework for CGM. In this paper, we propose a novel method that compresses CGM time series into a compact format which can be reconstructed on demand. Our approach preserves clinically relevant glucose dynamics while achieving high compression ratios. The encoded format is lightweight for storage and transmission, easing the burden on FHIR-based systems and enabling integration with blockchain or other secure platforms. At the same time, the decoded data can be restored for fine-grained analysis, and the encoding is usable in AI models.

# Methods

We used two complementary CGM datasets - synthetic data generated via a Conditional Generative Adversarial Network (CGAN) [18,19] and real-world CGM measurements from a randomized crossover trial [20] - to develop and validate two encoding methods based on peaks/nadirs information and peaks/nadirs + support points, and compare with uniform downsampling. The encoding reduces each profile to a small set of values, and reconstruction of the CGM signal uses interpolation. We compare compression ratios, reconstruction error (mean absolute error, MAE), and established CGM glycemic metrics between original and decoded signals. Statistical analyses assess whether encoding preserves clinically relevant glucose features. An overview of the methodology is illustrated in *Figure 1*.

## Data Sources

*Discovery Data*

We used the publicly available synthetic CGM dataset [18,19] generated by a CGAN trained on real CGM time series from healthy individuals and patients with type 1 diabetes across four $HbA_1c$ groups (<6.5%, 6.5–<7%, 7–<8%, ≥8%). In brief, the CGAN architecture was based on conditional on class label ($HbA_1c$ group), with generator and discriminator networks following the Mirza & Osindero formulation [21]. The dataset included 40,000, 24-hour profiles with a sample frequency of 288 measurements per day.

*Clinical Validation Data*

For validation, we used CGM baseline recordings from a randomized, open-label, crossover trial in patients with chronic pancreatitis and insulin-treated diabetes (n=30; mean age 64.4 ± 8.8 years; 75.9% male), each undergoing 20 days of CGM (Dexcome G6, sampling rate 5-min intervals) for the baseline [20].

## Encoding and Decoding Algorithms

We compared three encoding approaches for each CGM profile, in brief:

1. Peaks & Nadirs (PN) Only: Identify local maxima ("peaks") and minima ("nadirs") by sign changes in the discrete derivative of the glucose time series, subject to a minimum prominence threshold to avoid noise

2. Peaks, Nadirs, and Support Points (PN+)**:** This method extends the PN approach by incorporating additional "support" points between peaks and nadirs. These points were selected using a greedy search strategy to better capture the overall shape of the glucose trajectory.

3. Uniform Downsampling: As a baseline comparison, we applied uniform downsampling by removing every *x*-th sample to simulate a reduced sampling rate.

For each signal landmark (peak, nadir, or support point), we record its timestamp and glucose value.

*Signal reconstruction*

Decoded profiles were reconstructed by Piecewise Cubic Hermite Interpolating Polynomial (PCHIP) [22] using the identified landmarks, yielding a time series equivalent to the original 5-min sampling rate. This simple method preserves key excursion patterns while minimizing computational load.

*Peak Detection in Glucose Time Series*

To identify the significant excursions in glucose levels, we employed a peak detection algorithm. Glucose values were treated as a one-dimensional time series $g(t)$, where $t$ represents the corresponding timestamps of glucose measurements.

Local maxima were identified with a minimum peak prominence threshold of 15 mg/dL to ensure physiological relevance and to reduce the influence of minor fluctuations and noise. In this context, prominence quantifies how much a peak stands out due to its height and separation from neighboring valleys. Specifically, the prominence of each peak is computed as the vertical distance between the peak and the lowest point in the signal, separating it from a higher neighboring peak.

The detection criterion can be formally described as follows: for each candidate peak at time $t_i$, it is retained only if:

$$g(t_i) - max(min_{t_k<t_i} g_k^{\downarrow}, min_{t_k>t_i} g_k^{\downarrow}) \geq 15 \; mg/dL$$

where $g_k^{\downarrow}$ represents local valleys in the signal. This approach yields a set of clinically relevant glucose peaks. For nadir detections, the same procedure was repeated with an inverted signal.

*Adding additional support points*

To efficiently approximate CGM profiles using the limited number of representative landmarks, we implemented a greedy optimization strategy for selecting support points. The goal was to identify a sparse subset of additional points that allows accurate reconstruction of the original signal using PCHIP.

Given an initial subset of $k$ known landmark points $K_0 = \{(t_{i1}, g_{i1}), ..., (t_{ik}, g_{ik})\}$, where $i_j \in \{1,..,N\}$, the goal is to iteratively augment this set to a total of $n$ points by selecting additional time points that minimize reconstruction error. The upper bound for $n$ is constrained by the desired compression ratio.

At each iteration $j$, the algorithm evaluates all remaining candidate time points $t_m \notin K_j$ and selects the one that, when added, minimizes the L2-norm of the reconstruction error:

$$E_t(t_m) = \left|\left|g(t) - \hat{g}(K_j; t_m)\right|\right|_2$$

Where $\hat{g}(K_j; t_m)$ is the interpolated glucose signal using PCHIP over the updated key set $K_{j+1} \cup \{(t_m, g(t_m))\}$. The optimal new point is selected as:

$$t^* = \underset{t_m \notin K_j}{\arg\min} \; E(t_m)$$

This process repeats until $|K_j|$ reached the maximum number of desired landmark points. The final reconstructed signal $\hat{g}(t)$ is generated via: $\hat{g}(t) = PCHIP(K_j, t)$.

*Figure 2* illustrates the selected landmark points extracted from the CGM profiles, including both PN points as well as the additional support points used in the extended PN+ approach. The corresponding reconstructed signals for both methods are shown. Notably, the inclusion of support points in the PN+ approach improves the alignment of the reconstructed signal with the original glucose profile, particularly in regions between peaks and nadirs where the rate of change is non-uniform.

**Assessments**

To assess the compression of the approaches (PN/PN+/downsampling) we calculated the compression ratio between the uncompressed profiles and the compressed profiles:

$$Compression\ ratio = \frac{uncompressed}{compressed}$$

Furthermore, we calculated the MAE between original and decoded glucose profiles metrics:

$$MAE = \frac{1}{N}\sum_{i=1}^{n}\left|metric(P_i)_{orginal} - metric(P_i)_{reconstructed}\right|$$

To assess if the encoding profiles preserve clinically relevant glucose features, we compute standard consensus CGM metrics and glycemic variability metrics [23–25] on both original and decoded profiles. The MEA and coefficient of determination ($R^2$) were calculated for each metric between the original and decoded profiles. The standard consensus CGM metric included were mean glucose, standard deviation (SD) of glucose, coefficient of variation (CV), time-in-range (TIR, 70-180 mg/dL), time-in-tight-range (TITR, 70-140 mg/dL), time-below-range (TBR, <70 mg/dL), time-below range level 1 (TBR1, 54-70 mg/dL), time-below range level 2 (TBR2, <54 mg/dL), time-above-range (TAR, >180 mg/dL), time-above-range level 1 (TAR1, 180-250 mg/dL), time-above-range level 2 (TAR2, >250 mg/dL), and Mean Amplitude of Glycemic Excursions (MAGE) [26].

Analyses were performed in MATLAB vR2021b (MATLAB, The MathWorks Inc., Natick, Massachusetts, United States). For computing CGM metrics, we utilized an open-source tool, Quantification of Continuous Glucose Monitoring (QoCGM), designed for CGM data analysis using the MATLAB environment [27].

# Results

*Discovery Data*

As shown in *Figure 3*, the methods were compared across various settings - downsampling rate, peak prominence threshold, and the allowed number of total support points - to evaluate performance under different compression rates. Among the methods, PN+ consistently achieved the highest average $R^2$, while downsampling yielded the lowest performance at each comparable compression ratio. Furthermore, the impact of varying peak prominence thresholds on the PN and PN+ approaches is illustrated in *Figure 4*. As expected, PN+ resulted in lower compression at each threshold level due to the inclusion of additional support points. However, when comparing the methods at equivalent compression ratios, PN+ exhibited substantially lower error than PN. This suggests that PN+ is a more effective approach for capturing clinically relevant physiological signal patterns. The encoding / decoding combined took on average of 0.13 seconds per 24-hour profile on a PC laptop (11th Gen Intel(R) Core(TM) i7-11850H @ 2.50GHz, 32,0 GB RAM).

A direct comparison of CGM-derived metrics from the reconstructed signals and the original signals is presented in *Table 1*, using a peak prominence threshold of 15 mg/dL and a minimum compression ratio of 13 (corresponding to 22 landmark points out of the original 288). At this compression level, PN+ consistently achieved higher $R^2$ values across most CGM-derived metrics compared to downsampling. More notably, the MAE for each metric was substantially lower with PN+, with an average reduction factor of 3.6 (0.77 vs. 2.75). This improvement was especially pronounced for MAGE, which captures large glucose excursions — an area where downsampling introduced significant estimation errors as illustrated in Figure 5 by an Bland-Altman analysis plot.

Overall, PN+ demonstrates strong alignment between metrics derived from reconstructed and original signals. However, the estimation of TBR2, representing rare low-glucose episodes, remains sensitive to the encoding and decoding process due to the sparse occurrence of such events over the typical day.

*Clinical Validation Data*

As shown in *Table 2*, the validation dataset yielded overall performance trends similar to those observed in the discovery dataset. However, the $R^2$ and MAE values were slightly worse across all methods, likely reflecting the increased complexity and variability inherent in real-world patient data. Notably, the validation cohort included individuals with diabetes associated with chronic pancreatitis, a diabetes subtype known for markedly impaired glucose regulation and high glycemic variability due to deficiencies in insulin and glucagon secretion, along with variable insulin sensitivity [28]. Despite this added complexity, PN+ continued to outperform downsampling at equivalent compression ratios, achieving the lowest error and highest $R^2$ among the compared methods.

## Discussion

The proposed novel encoding algorithm PN+ demonstrated advantages for CGM data compression by explicitly selecting physiologically meaningful points. By anchoring the compressed signal on glucose peaks and nadirs (and additional support points via a greedy strategy), PN+ yields substantially lower reconstruction error and higher explained variance ($R^2$) across CGM-derived metrics than uniform downsampling at equivalent compression ratios. This improvement is especially notable for glycemic variability measures. In particular, PN+ better preserves MAGE, a gold-standard CGM metric for assessing large glycemic excursions [29]. Because MAGE and similar variability indices depend on accurately capturing excursions, PN+'s targeted point selection translates directly to better metric fidelity. Furthermore, reconstructing the compressed signal with PCHIP maintains the original shape and smoothness of glucose trends [30], avoiding the overshoot or oscillation that might arise with simpler interpolation. Overall, the proposed algorithm was able to reconstruct the CGM signals, preserving key clinically relevant features of the curves from a substantially decompressed version.

To date, relatively few studies have addressed the compression or encoding of CGM profiles. However, Kovatchev *et al* [31] showed using a multistep machine-learning procedure, how it is possible to reconstruct a virtual CGM profile from the original sparse data (7-point blood glucose profiles) - while preserving clinically relevant information [32].Notably, Igbe and Kovatchev [15] recently introduced a novel approach that encodes daily CGM profiles into symbolic representations - referred to as CGM strings and CGM texts - which preserve key clinical metrics while achieving data compression. Their analysis demonstrated that a nine-character encoding corresponding to an approximate compression ratio of 10:1 could retain information on CV with an $R^2$ of up to 0.93. The method presented in this study builds on and extends these findings by enabling the preservation of a broader range of CGM-derived metrics across multiple compression levels. Moreover, our approach offers the additional advantage of reconstructing the original signal with high precision, supporting both clinical interpretability and downstream analytical use.

The proposed approach is fundamentally a content-based compression strategy and could be applied to other periodic or episodic biomedical signals. For example, electrocardiogram (ECG) waveforms hinge on distinct QRS-complex peaks for heartbeats, and compression methods that preserve those peaks can maintain diagnostic integrity. Indeed, prior work has shown that a lossy ECG compression scheme preserved the main features of the ECG morphology (notably the QRS complexes) despite a high compression ratio of 4.5 [33]. Similarly, photoplethysmographic (PPG) waveforms depend on capturing the systolic pulse peaks; a PN+-like method could retain these pulse peaks as key support points. In both ECG and PPG contexts, the idea is the same as for CGM: allocate sampling "budget" to physiologically salient events (heartbeats or pulses) and interpolate the rest. Thus, PN+ is broadly relevant to any biosignal where preserving the height and timing of key peaks and nadirs is critical for clinical or functional interpretation. Future work should explore the potential applicability of the proposed methodology to other biological signals. Moreover, future work should also investigate the clinical impact on decision-making based on reconstructed signal-data.

**Strengths and Limitations**

This study is based on a large synthetic dataset of CGM profiles derived from a heterogeneous population of individuals with diabetes. Furthermore, the proposed methods were externally validated using data from a distinct patient group with diabetes secondary to chronic pancreatitis. These findings could support the generalizability of the results to a broader population of individuals with diabetes.
However, the proposed PN+ method has some limitations typical of lossy, feature-based compression approaches. One concern is rare event sensitivity: very short or abrupt excursions (such as sudden severe hypoglycemia events) might not always produce easily identified peaks or may occur between selected support points. In practice, we observed that extremely brief dips can be slightly underrepresented by PN+ unless thresholds are tuned to catch them. Relatedly, the performance gain of PN+ depends on the presence of signal structure. If a CGM trace is unusually flat or noisy (e.g. due to sensor artifacts), the greedy peak-nadir selection may yield fewer benefits, and overall reconstruction error for all methods can rise. Indeed, in our heterogeneous validation cohort, all compression methods saw modestly reduced accuracy, though PN+ remained superior; this suggests that real-world variability poses challenges even to structured approaches. Lastly, PN+ currently uses fixed rules (greedy selection and threshold-based peaks/nadirs identification), so it may require adaptation (e.g. dynamic thresholds or rule changes) when applied to data with very different characteristics.

**Conclusion**

The proposed PN+ method produces a compact CGM representation that retains critical glycemic dynamics while discarding redundant portions of the signal. The CGM signal can be reconstructed with high precision from the encoding representation. The application of the method to other biomedical signals needs further investigation.

**Acknowledgments**

None to report

**Funding received**

None to report

**Disclosures**

SLC has received research funding from i-SENS Inc., holds shares in Novo Nordisk A/S, and has received consultancy fees from Roche Diagnostics and Medicus Engineering.

**Ethical Statement**

All procedures were performed in compliance with relevant laws and institutional guidelines and have been approved by the appropriate institutional committee(s). For the validation data, the North Denmark Region Committee on Health Research Ethics approved the protocol (N-20210064), and the study adhered to the Declaration of Helsinki and Good Clinical Practice. Written informed consent was obtained before enrollment.

**Code Availability**

The code used to generate the results reported in this study will be made publicly available in a GitHub repository (https://github.com/simcich/CGM_Data_Compression_Reconstruction) upon acceptance of this manuscript for publication. Access to the repository will enable other researchers to reproduce the analyses and facilitate further development of the methods described.

**Data Availability**

The discovery dataset used in this study is openly available at [Mendeley Data](). The clinical validation dataset contains sensitive personal information and, therefore, access is restricted; requests for access may be considered under appropriate data use agreements and ethical approval.

# References

1   Battelino T, Lalic N, Hussain S, *et al.* The use of continuous glucose monitoring in people living with obesity, intermediate hyperglycemia or type 2 diabetes. *Diabetes Res Clin Pract*. 2025;223:112111. doi: 10.1016/J.DIABRES.2025.112111

2   Klonoff DC, Ahn D, Drincic A. Continuous glucose monitoring: A review of the technology and clinical use. *Diabetes Res Clin Pract*. 2017;133:178–92. doi: 10.1016/J.DIABRES.2017.08.005

3   Bender C, Vestergaard P, Cichosz SL. The History, Evolution and Future of Continuous Glucose Monitoring (CGM). *Diabetology 2025, Vol 6, Page 17*. 2025;6:17. doi: 10.3390/DIABETOLOGY6030017

4   Woldaregay AZ, Årsand E, Walderhaug S, *et al.* Data-driven modeling and prediction of blood glucose dynamics: Machine learning applications in type 1 diabetes. *Artif Intell Med*. 2019;98:109–34. doi: 10.1016/j.artmed.2019.07.007

5   Kahkoska AR, Adair LA, Aiello AE, *et al.* Identification of clinically relevant dysglycemia phenotypes based on continuous glucose monitoring data from youth with type 1 diabetes and elevated hemoglobin A1c. *Pediatr Diabetes*. 2019;20:556–66. doi: 10.1111/PEDI.12856,

6   Ayers AT, Ho CN, Kerr D, *et al.* Artificial Intelligence to Diagnose Complications of Diabetes. *J Diabetes Sci Technol*. Published Online First: 13 September 2024.

7   Lebech Cichosz S, Kronborg ; Thomas, Laugesen E, *et al.* From Stability to Variability: Classification of Healthy Individuals, Prediabetes, and Type 2 Diabetes using Glycemic Variability Indices from Continuous Glucose Monitoring Data. *https://home.liebertpub.com/dia*. Published Online First: 8 August 2024. doi: 10.1089/DIA.2024.0226

8   Cichosz SL, Jensen MH, Olesen SS. Development and Validation of a Machine Learning Model to Predict Weekly Risk of Hypoglycemia in Patients with Type 1 Diabetes Based on Continuous Glucose Monitoring. *https://home.liebertpub.com/dia*. Published Online First: 12 January 2024. doi: 10.1089/DIA.2023.0532

9   Cichosz SL, Bender C. Development of Machine Learning Models for the Identification of Elevated Ketone Bodies During Hyperglycemia in Patients with Type 1 Diabetes. *https://home.liebertpub.com/dia*. Published Online First: 8 March 2024. doi: 10.1089/DIA.2023.0531

10  Lacy ME, Lee KE, Atac O, *et al.* Patterns and Trends in Continuous Glucose Monitoring Utilization Among Commercially Insured Individuals With Type 1 Diabetes: 2010–2013 to 2016–2019. *Clinical Diabetes*. 2024;42:388–97. doi: 10.2337/CD23-0051/745990/DIACLINCD230051.PDF

11  Jacobsson M, Seoane F, Abtahi F. The role of compression in large scale data transfer and storage of typical biomedical signals at hospitals. *Health Informatics J*. 2023;29. doi: 10.1177/14604582231213846,

12  Lal B, Gravina R, Spagnolo F, *et al.* Compressed Sensing Approach for Physiological Signals: A Review. *IEEE Sens J*. 2023;23:5513–34. doi: 10.1109/JSEN.2023.3243390

13  Nassra I, Capella J V. Data compression techniques in IoT-enabled wireless body sensor networks: A systematic literature review and research trends for QoS improvement. *Internet of Things*. 2023;23:100806. doi: 10.1016/J.IOT.2023.100806

14  Cichosz SL, Stausholm MN, Kronborg T, *et al.* How to Use Blockchain for Diabetes Health Care Data and Access Management: An Operational Concept. *J Diabetes Sci Technol*. 2018;13:248. doi: 10.1177/1932296818790281

15  Igbe T, Kovatchev B. Finding Optimal Alphabet for Encoding Daily Continuous Glucose Monitoring Time Series Into Compressed Text. *J Diabetes Sci Technol*. 2025;19322968251323910. doi: 10.1177/19322968251323913

16  Hua J, Chu B, Zou J, *et al.* ECG signal classification in wearable devices based on compressed domain. *PLoS One*. 2023;18:e0284008. doi: 10.1371/JOURNAL.PONE.0284008

17  Wu Z, Guo C. Deep learning and electrocardiography: systematic review of current techniques in cardiovascular disease diagnosis and management. *BioMedical Engineering OnLine 2025 24:1*. 2025;24:1–50. doi: 10.1186/S12938-025-01349-W

18  Cichosz S, Xylander A. Synthetic Continuous Glucose Monitoring (CGM) Signals. *Mendelay data*. Published Online First: 2021. doi: 10.17632/chd8hx65r4.1

19  Cichosz SL, Xylander AAP. A Conditional Generative Adversarial Network for Synthesis of Continuous Glucose Monitoring Signals: *https://doi.org/101177/19322968211014255*. Published Online First: 30 May 2021. doi: 10.1177/19322968211014255

20  Davidsen L, Cichosz SL, Stæhr PB, *et al.* Efficacy and safety of continuous glucose monitoring on glycaemic control in patients with chronic pancreatitis and insulin-treated diabetes: A randomised, open-label, crossover trial. *Diabetes Obes Metab*. 2025;27. doi: 10.1111/DOM.16356,

21  Mirza M, Osindero S. Conditional Generative Adversarial Nets. Published Online First: 6 November 2014.

22  Rabbath CA, Corriveau D. A comparison of piecewise cubic Hermite interpolating polynomials, cubic splines and piecewise linear functions for the approximation of projectile aerodynamics. *Defence Technology*. 2019;15:741–57. doi: 10.1016/J.DT.2019.07.016

23  Danne T, Nimri R, Battelino T, *et al.* International Consensus on Use of Continuous Glucose Monitoring. *Diabetes Care*. 2017;40:1631–40. doi: 10.2337/DC17-1600

24  Battelino T, Alexander CM, Amiel SA, *et al.* Continuous glucose monitoring and metrics for clinical trials: an international consensus statement. *Lancet Diabetes Endocrinol*. 2023;11:42–57. doi: 10.1016/S2213-8587(22)00319-9

25  Battelino T, Danne T, Bergenstal RM, *et al.* Clinical targets for continuous glucose monitoring data interpretation: Recommendations from the international consensus on time in range. *Diabetes Care*. 2019;42:1593–603. doi: 10.2337/dci19-0028

26  Service FJ, Molnar GD, Rosevear JW, *et al.* Mean Amplitude of Glycemic Excursions, a Measure of Diabetic Instability. *Diabetes*. 1970;19:644–55. doi: 10.2337/DIAB.19.9.644

27     Lebech Cichosz S, Hangaard ; Stine, Kronborg T, *et al.* From Data to Insights: A Tool for Comprehensive Quantification of Continuous Glucose Monitoring (QoCGM). *medRxiv*. 2025;2025.01.01.25319870. doi: 10.1101/2025.01.01.25319870

28     Davidsen L, Jensen MH, Kronborg T, *et al.* Increased Glycemic Variability in Patients With Chronic Pancreatitis and Diabetes Compared to Type 2 Diabetes. *J Diabetes Sci Technol*. 2025;19322968251356240. doi: 10.1177/19322968251356239

29     Vergès B, Pignol E, Rouland A, *et al.* Glycemic Variability Assessment with a 14-Day Continuous Glucose Monitoring System: When and How Long to Measure MAGE (Mean Amplitude of Glucose Excursion) for Optimal Reliability? *J Diabetes Sci Technol*. 2021;16:982. doi: 10.1177/1932296821992060

30     Lebech Cichosz S, Hangaard ; Stine, Kronborg T, *et al.* From Data to Insights: A Tool for Comprehensive Quantification of Continuous Glucose Monitoring (QoCGM). *medRxiv*. 2025;2025.01.01.25319870. doi: 10.1101/2025.01.01.25319870

31     Kovatchev BP, Lobo B, Fabris C, *et al.* The Virtual DCCT: Adding Continuous Glucose Monitoring to a Landmark Clinical Trial for Prediction of Microvascular Complications. *Diabetes Technol Ther*. 2025;27:2025. doi: 10.1089/DIA.2024.0404,

32     Lobo B, Kanapka L, Kovatchev BP, *et al.* The Association of Time-in-Range and Time-in-Tight-Range with Retinopathy Progression in the Virtual Diabetes Control and Complications Trial Continuous Glucose Monitoring Dataset. *Diabetes Technol Ther*. Published Online First: 2025. doi: 10.1089/DIA.2025.0033,

33     Elgendi M, Mohamed A, Ward R. Efficient ECG Compression and QRS Detection for E-Health Applications. *Scientific Reports 2017 7:1*. 2017;7:1–16. doi: 10.1038/s41598-017-00540-x

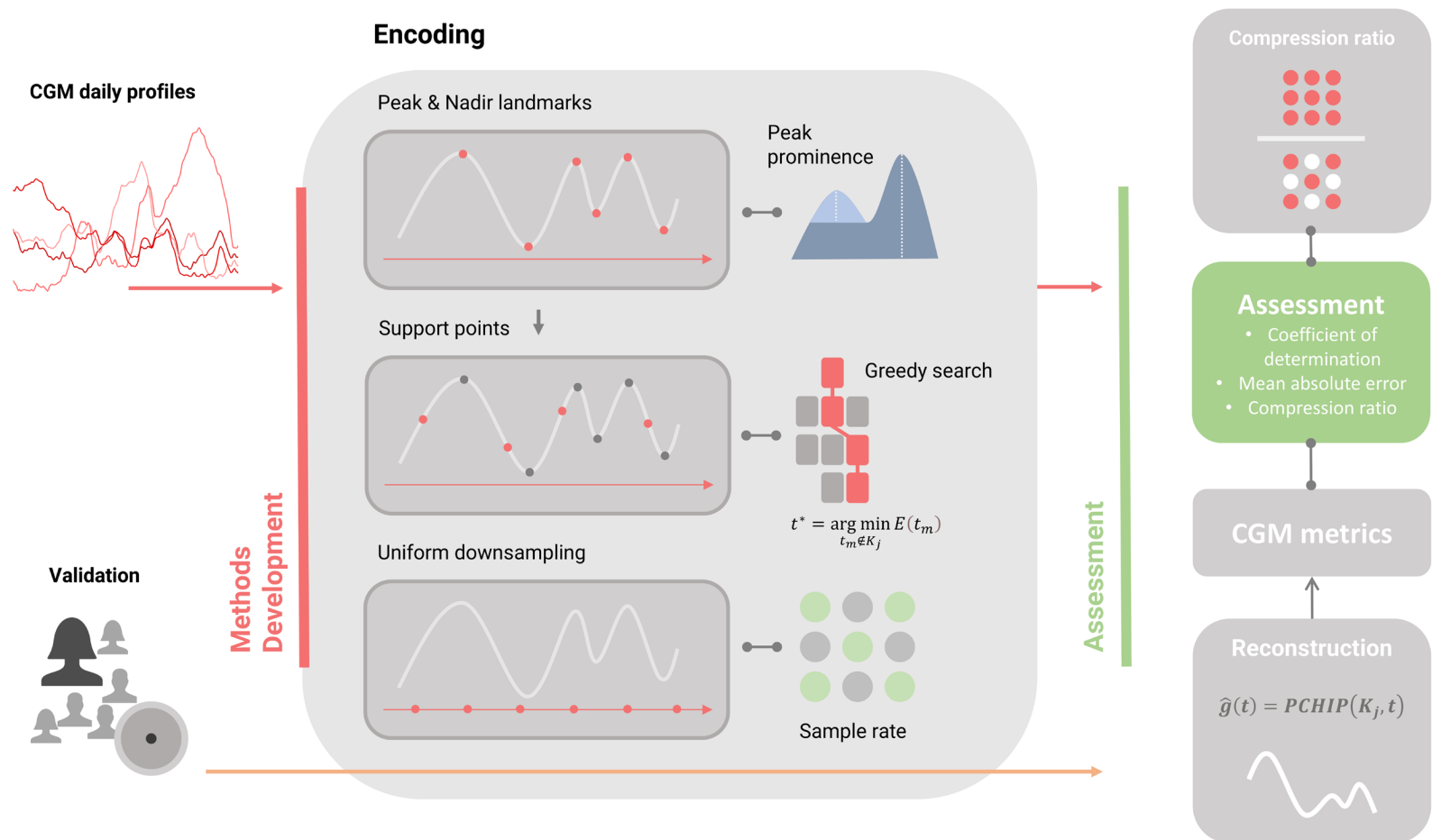

Figure 1 - Overview of the methodological framework. Continuous glucose monitoring (CGM) daily profiles were encoded using three approaches: (i) peak and nadir landmarks identified by peak prominence, (ii) support points selected through a greedy search procedure, and (iii) uniform downsampling at a predefined sampling rate. The encoded signals were subsequently reconstructed using piecewise cubic Hermite interpolation (PCHIP). Reconstruction performance was evaluated based on the coefficient of determination ($R^2$), mean absolute error, and compression ratio.

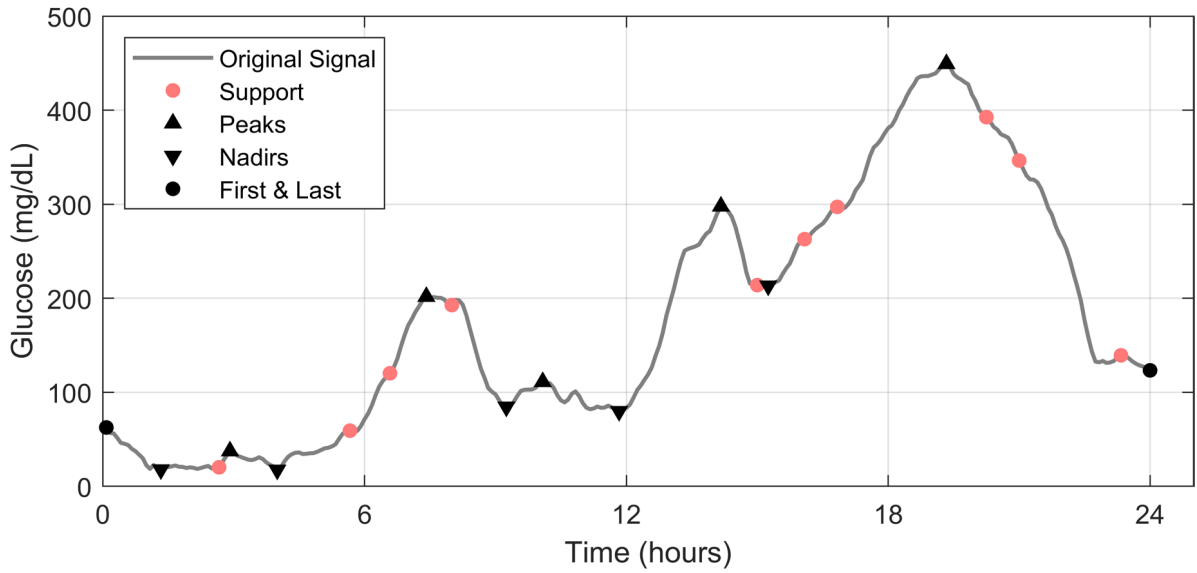
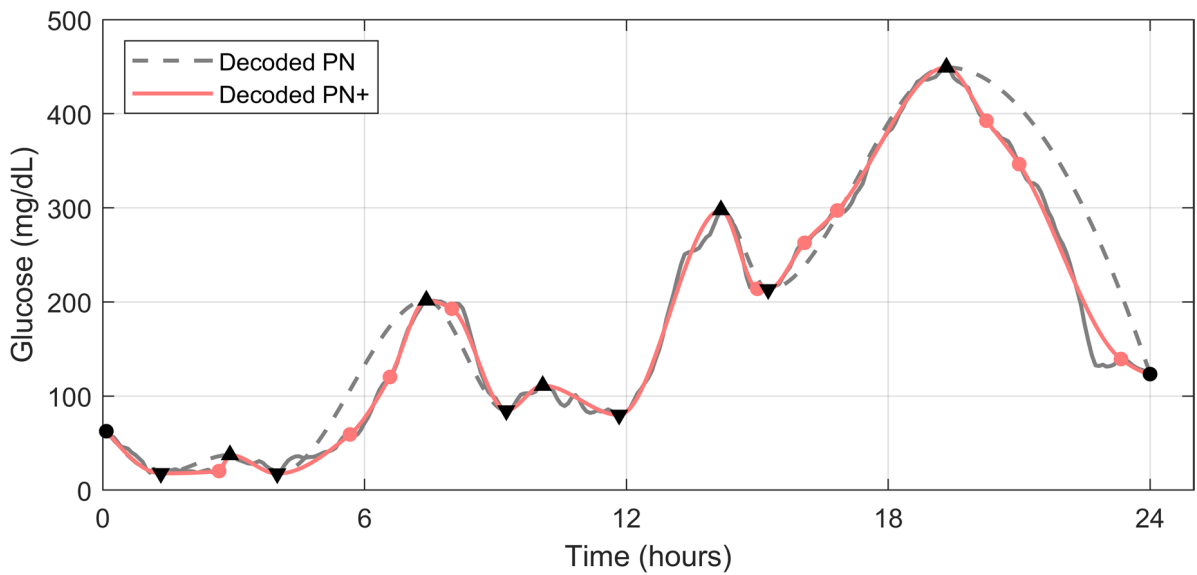

Figure 2 – The top panel shows an example of a 24-hour CGM profile (original signal) with the peak, nadir, and support points identified by the algorithm. The bottom panel shows the reconstructed signals using only the peak/nadir points (PN) and using the method with additional support points (PN+).

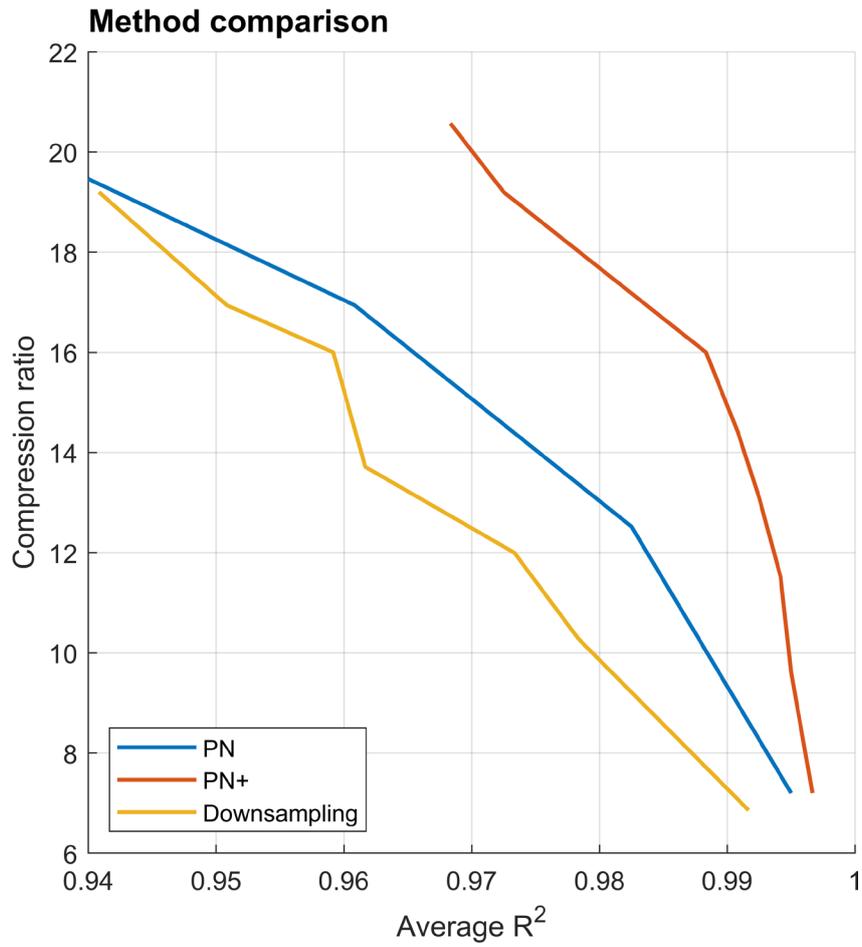

Figure 3 – Compression ratio plotted against the average $R^2$ values for the three methods across all glycemic metrics derived from the reconstructed CGM signals.

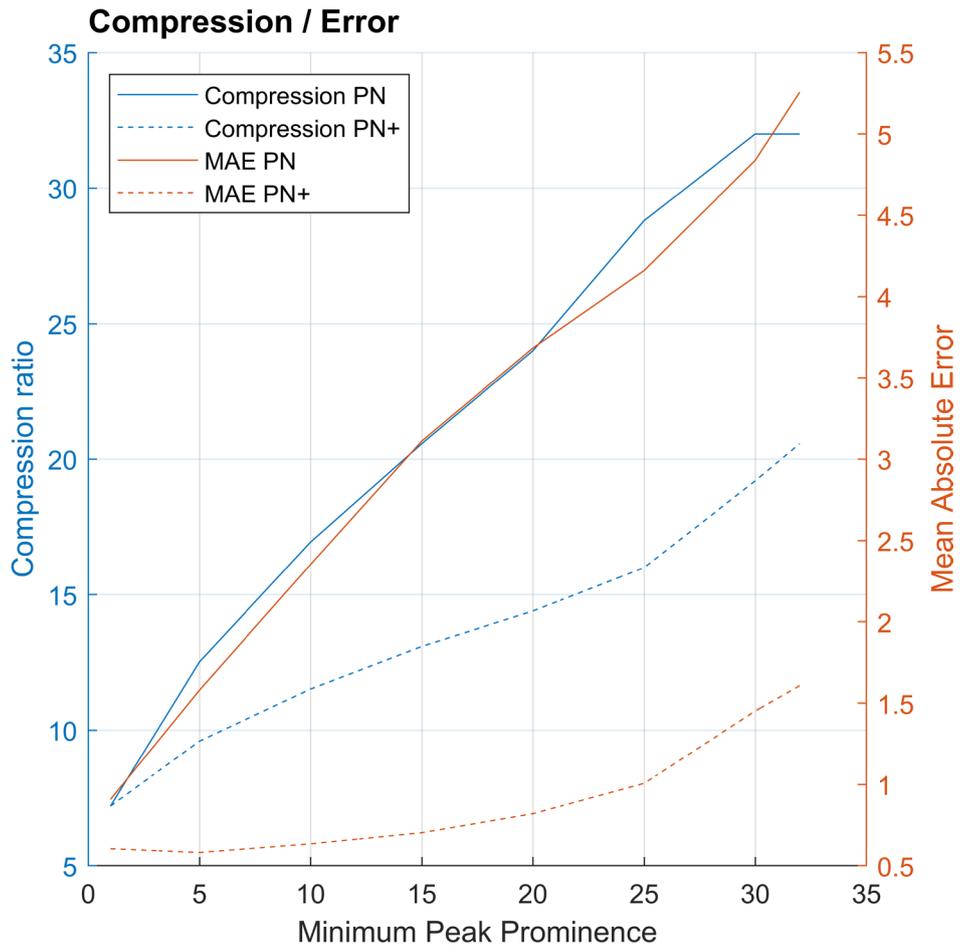

Figure 4 - impact of varying peak prominence thresholds on the PN and PN+ approaches mean absolute error.

# Bland-Altman | Analysis

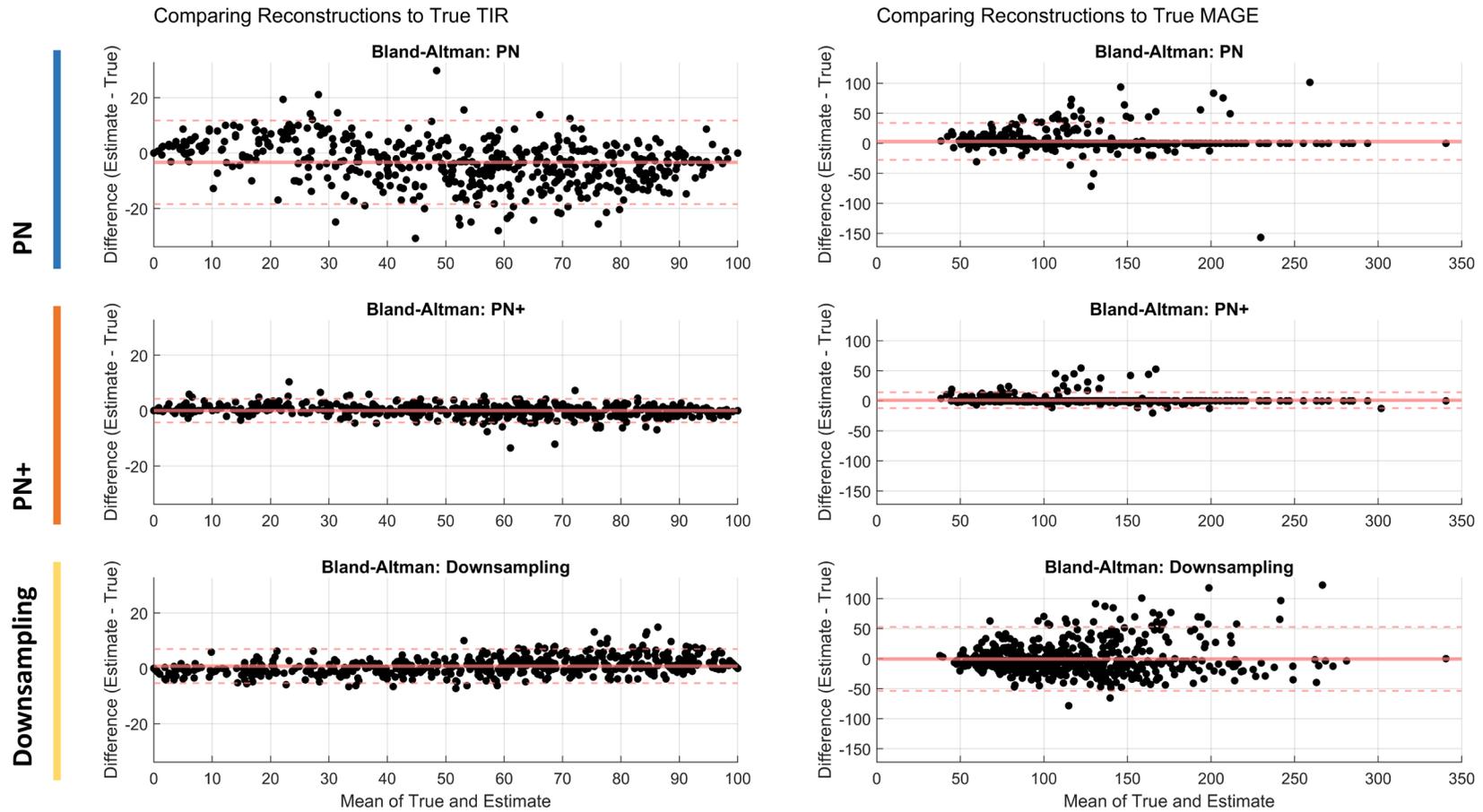

Figure 5 – Bland-Altman analysis plots comparing the three methods (PN, PN+ and Downsampling) for reconstruction error estimates of clinical metrics (TIR and MAGE) on clinical validation data.

**Discovery data**

|  | PN | | PN+ | | Downsampling | |
|---|---|---|---|---|---|---|
|  | $R^2$ | MAE | $R^2$ | MAE | $R^2$ | MAE |
| Mean glucose, mg/dl | 0.98 | 5.24 | 1 | 0.64 | 1 | 0.61 |
| CV, ratio | 0.97 | 2.01 | 1 | 0.50 | 0.99 | 2.02 |
| SD glucose, mg/dl | 0.98 | 3.3 | 1 | 0.65 | 1 | 2.91 |
| TIR, % | 0.97 | 3.36 | 1 | 1.14 | 0.99 | 2.12 |
| TITR, % | 0.97 | 3.64 | 1 | 1.36 | 0.99 | 2.36 |
| TBR, % | 0.90 | 1.52 | 0.98 | 0.66 | 0.97 | 1.1 |
| TBR1, % | 0.75 | 1.36 | 0.92 | 0.82 | 0.89 | 1.11 |
| TBR2, % | 0.90 | 0.66 | 0.98 | 0.32 | 0.96 | 0.47 |
| TAR, % | 0.97 | 2.79 | 1 | 0.70 | 0.99 | 1.38 |
| TAR1, % | 0.92 | 2.76 | 0.99 | 0.89 | 0.97 | 1.87 |
| TAR2, % | 0.96 | 1.89 | 1 | 0.40 | 0.99 | 0.98 |
| MAGE, mg/dl | 0.97 | 3.71 | 0.99 | 1.1 | 0.90 | 16.13 |
| Overall mean | 0.94 | 2.69 | 0.99 | 0.77 | 0.98 | 2.75 |
| Compression ratio | 20.6 | | 13.1 | | 13.1 | |

Table 1. Overall performance ($R^2$ and MAE) and compression ratio of Peaks & Nadir (PN) encoding, with and without support points (PN+), and downsampling for each CGM metric in the discovery dataset (n = 40,000).

**Validation data**

|  | PN | | PN+ | | Downsampling | |
|---|---|---|---|---|---|---|
|  | $R^2$ | MAE | $R^2$ | MAE | $R^2$ | MAE |
| Mean glucose, mg/dl | 0.96 | 8.13 | 1 | 0.99 | 1 | 1.4 |
| CV, ratio | 0.87 | 2.68 | 0.99 | 0.69 | 0.97 | 1.68 |
| SD glucose, mg/dl | 0.89 | 5.16 | 0.99 | 1.15 | 0.98 | 2.9 |
| TIR, % | 0.93 | 6.26 | 0.99 | 1.43 | 0.99 | 2.32 |
| TITR, % | 0.89 | 5.84 | 0.99 | 1.63 | 0.98 | 2.34 |
| TBR, % | 0.75 | 0.95 | 0.95 | 0.33 | 0.90 | 0.49 |
| TBR1, % | 0.60 | 0.80 | 0.92 | 0.31 | 0.91 | 0.39 |
| TBR2, % | 0.75 | 0.41 | 0.84 | 0.16 | 0.56 | 0.16 |
| TAR, % | 0.94 | 5.65 | 1 | 1.24 | 0.99 | 2.01 |
| TAR1, % | 0.79 | 5.83 | 0.98 | 1.57 | 0.96 | 2.89 |
| TAR2, % | 0.95 | 3.56 | 1 | 0.81 | 0.99 | 1.5 |
| MAGE, mg/dl | 0.90 | 5.79 | 0.98 | 1.86 | 0.72 | 20.2 |
| Overall mean | 0.85 | 4.25 | 0.97 | 1.01 | 0.91 | 3.19 |
| Compression ratio | 24.1 | | 13.1 | | 13.1 | |

Table 2. Overall performance ($R^2$ and MAE) and compression ratio of Peaks & Nadir (PN) encoding, with and without support points (PN+), and downsampling for each CGM metric in the clinical validation dataset (n = 30).